\documentstyle[preprint,aps,pre]{revtex} 
 
\begin{document} 
 
%\draft 
 
\title{Anomalous diffusion with absorption: Exact time-dependent solutions} 
 
\author{German Drazer$^1$ 
\thanks{Electronic address:gdrazer@tron.fi.uba.ar}, 
Horacio S. Wio $^2$\thanks{Electronic address: wio@cab.cnea.gov.ar, 
\newline http://www.cab.cnea.gov.ar/Cab/invbasica/FisEstad/estadis.htm} 
and  Constantino Tsallis $^3$  
\thanks{Electronic address: tsallis@cbpf.br}}  
 
\address{$^1$ Grupo de Medios Porosos. Facultad de Ingenier\'{\i}a. Universidad 
de Buenos Aires. \\ Paseo Col\'on 850. CP 1063, Argentina} 
\address{$^2$ Centro At\'omico Bariloche (CNEA) and Instituto Balseiro (CNEA and UNC) \\ 
 8400--San Carlos de Bariloche, Argentina.} 
\address{$^3$ Centro Brasileiro de Pesquisas F\'{\i}sicas, Rua Xavier Sigaud 150, 22290-180 
\\ Rio de Janeiro, Brazil\\
and Department of Physics, University of North Texas, Denton, Texas 76203, USA.} 
\date{\today} 

\maketitle 
 
\begin{abstract} 
Recently, analytical solutions of a nonlinear Fokker-Planck equation  
describing anomalous diffusion with an external linear force were found  
using a non extensive thermostatistical Ansatz. We have extended these   
solutions to the case when an homogeneous absorption process is also 
present. Some peculiar aspects of the interrelation between the 
deterministic force, the nonlinear diffusion and the absorption process  
are discussed. 
\end{abstract} 
 
 
%} 
 
\section{Introduction}

The ubiquity of the anomalous diffusion phenomenon in nature has attracted the  
interest of researchers from both the theoretical and experimental point of view.  
Anomalous diffusion has been found in transport of fluids in porous media and  
surface growth \cite{spohn}, in NMR relaxometry of liquids in porous glasses  
\cite{bychuk}, in a two dimensional fluid flow \cite{solo}, to name just a few  
among the large variety of physical phenomena where it is present. A related  
aspect is the case of density dependent diffusivities as found in some biological  
systems \cite{MURR}, in polymers \cite{polimeros}, and hydrogen diffusion in metals  
\cite{fukai} (see also \cite{strier}).  
 
Some recent papers have investigated a class of {\it nonlinear} generalizations  
diffusion and Fokker--Planck equations \cite{spohn,inic3,inic4,inic1}, as a model  
of correlated anomalous diffusion. Some of those studies were based in a nonextensive thermodynamical formalism \cite{tsa}.  
Particularly in Ref. \cite{inic1} exact solutions for the nonlinear Fokker--Planck  
equation subject to a linear force have been found. Here we want to show how these  
solutions can be extended to the case where an absorption process is also present.  
 
We start recalling the ``full" anomalous diffusion equation or  
``nonlinear" Fokker-Planck equation solved in Ref. \cite{inic1} 
\begin{equation} 
\label{general} 
{\partial \over \partial t} [p(x,t)]^{\mu} =  
- {\partial \over \partial x} \{F(x)[p(x,t)]^{\mu}\}  
+ D {\partial^2 \over \partial x^2} [p(x,t)]^{\nu}.  
\end{equation} 
When $F(x)=0$, Eq. (\ref{general}) can be interpreted as a diffusion equation for  
$\Phi(x,t) = [p(x,t)]^{\mu}$, where the diffusivity depends on $\Phi$  
\begin{eqnarray} 
\label{diffusionefectiva} 
{\partial \over \partial t} \Phi = {\partial ^2 \over \partial x^2} 
(D(\Phi) \Phi) \\ 
D(\Phi) = \Phi^{{\nu \over \mu} -1}.  
\end{eqnarray}  
There are several real situations where this power-law dependence of the  
diffusivity is found. It occurs in the flow of gases through 
porous media ($\nu/\mu \geq 2$ \cite{muzkat}), the flow of water 
in unsaturated soils ($\nu/\mu = 5$ \cite{wiest}), the simultaneous diffusion  
and adsorption in porous samples where the adsorption isotherm 
is of power-law type \cite{crank} ($\nu/\mu \geq 1$ for Freundlich 
type of adsorption isotherm \cite{crank,concentration}). Clearly, in those 
cases, the diffusivity vanishes (diverges) for $\Phi =0$ when 
$\nu / \mu > 1$ ($ \nu / \mu < 1$).  
It is worth to remember that Eq. (\ref{general}) corresponds to the so called  
``Porous Media Equation" when $\mu = 1$ \cite{inic4,pormedeq}.  
There are a large number of situations where the interest of describing anomalous  
diffusion plus absorption are of relevance. Notably, those related with diffusion  
of some (reactive) substance in gaseous phase through a porous media or a  
membrane, that react and can be adsorbed in sites inside the pore \cite{ref1}.  
 
Our interest here is to solve the same Eq. (\ref{general}) but including now  
terms that describe some kind of absorption process. A general form of  
such an equation is  
\begin{equation} 
\label{completa} 
{\partial \over \partial t} [p(x,t)]^{\mu} =  
- {\partial \over \partial x} \{F(x)[p(x,t)]^{\mu}\}  
+ D {\partial^2 \over \partial x^2} [p(x,t)]^{\nu}  
- \alpha [p(x,t)]^{\mu '},  
\end{equation} 
where $\alpha$ plays the role of an absorption rate (and becomes the usual one for 
$\mu ' = \mu $). 
The presence of reaction terms like the one in Eq. (\ref{completa}) (with  
$\alpha \neq 0$ and  $\mu ' \neq 0$) is not at all unexpected considering  
the large amount of work on the problematic of diffusion--limited reactions.  
Among the diversity of systems that have been studied we only recall here  
the so called one species coagulation, that is: $A+A \to 0$ or $m\,A \to l\,A$  
(with $m > l$), that have been associated, among others, with catalytic processes  
in regular, heterogeneous or disordered systems \cite{ejemplo}. 
The reaction term may account, in the case $\mu = \mu'$,  
for an irreversible first-order reaction of the transported substance so that 
the rate of removal is $\alpha\,C$ \cite{crank}. This extra term also appears when   
 a tracer undergoing radioactive decay is transported  
through a porous medium, where $\alpha$ is the reciprocal of the tracer's mean lifetime 
\cite{bear}, as well as in  heat flow involving heat production at a rate which is  
a linear function of the temperature \cite{carslaw}. Finally, in   
solute transport through adsorbent samples the adsorption rate, at small solute concentration,  
is usually proportional to the concentration in solution and  
Eq. (\ref{completa}) applies.  
 
In \cite{inic1} it has been shown that $p_q(x,t)$, the solution of   
Eq. (\ref{general}), for a linear force $F(x)$ has the form  
\begin{equation} 
\label{sol1} 
p_q(x,t)={\{1-\beta(t)(1-q)[x-x_M(t)]^2\}^{1/(1-q)} \over Z_q(t)},  
\end{equation} 
where $q=1+\mu-\nu$, $\beta(t)$ depends on the width of the distribution,  
$x_M(t)$ is the average of the coordinate and  $Z_q(t)$ is a normalization  
factor. All of them depend on the diffusion parameter $D$ as well as on  
$\mu , \nu$ and the force (see \cite{inic1} for details).  
 
For completeness, as well as for reference, we write Eq. (\ref{general})  
without the ``drift" $F(x)$ as we will refer to the solutions of such an  
equation in the following sections 
\begin{equation} 
\label{simple} 
{\partial \over \partial t} [p(x,t)]^{\mu} =  
D {\partial^2 \over \partial x^2} [p(x,t)]^{\nu}.  
\end{equation} 
According to the results from Ref. \cite{inic1} we can write that the  
solution $p_q^{(0)}(x,t)$ of Eq. (\ref{simple})  
has the form given in Eq. (\ref{sol1}), with  
\begin{eqnarray} 
\label{sol1c} 
x_M(t) &=& x_0 \nonumber \\ 
Z_q(t) &=& \left( \frac{2 \nu}{\mu} (\nu + \mu) \pi D t \right)^{\frac{1}{\nu +\mu}}\nonumber \\ 
\beta(t) &=& \pi \left( \frac{2 \nu}{\mu} (\nu + \mu) \pi D t  
\right)^{-\frac{2 \mu}{\nu +\mu}},  
\end{eqnarray} 
where, we have used the relation $\beta (0) Z_q(0)^{2 \mu} = \pi$, that shall   
be fulfilled if we want to have a $\delta$-like initial condition.  
 
In the present work we intent to analyze the specific case of a linear drift,  
namely $F(x)=k_1-k_2 x$. This case, where the potential is harmonic (a typical  
approximation), is the simple nontrivial one where analytic solutions can be  
obtained just by means of changing the 
variables to suitable ones, namely a simple extension of the well known 
Boltzmann Transformation \cite{boltz}.  
 
In section \ref{constant} we start considering the simple case $\mu=\mu'$. We analize 
first the case of constant external force ($k_2=0$). 
In this case we can  
firstly reduce Eq. (\ref{completa})  by proposing a solution of the form 
\begin{equation} 
\label{reduction} 
p(x,t) = e^{-{\alpha \over \mu} t}~\hat{p}(x,t),  
\end{equation} 
that yields an equation for $\hat{p}(x,t)$ given by 
\begin{eqnarray} 
\label{general2} 
{\partial \over \partial t} [\hat{p}(x,t)]^{\mu} &=&  
- {\partial \over \partial x} \{F(x)[\hat{p}(x,t)]^{\mu}\}  
\\ \nonumber 
&+& D~e^{\alpha(1-\nu/\mu) t} {\partial^2 \over \partial x^2} [\hat{p}(x,t)]^{\nu}.  
\end{eqnarray} 
Although this reduction to a {\it nonlinear} Fokker-Planck equation like in  
Eq. (\ref{general}) looks to be always possible (when $k_2=0$), for the case of  
a linear force (when $k_2 \neq 0$) such a reduction is not possible to make in  
a simple way and we will need a more general treatment. 
 
The linear force situation, tightly  
related to the so called Uhlenbeck-Ornstein process ($k_1=0;k_2 \neq 0$), is  
treated in section \ref{sec-general}, while in section \ref{sec-mu} we discuss  
the most general case, that is when  $\mu \neq \mu'$. In the last section we make  
some final remarks.

\section{Solution for a constant force} 
\label{constant} 
 
As indicated above, here we consider the case of a constant force, that is $F(x)=k_1$.  
Equation (\ref{general2}) can be further reduced making the following change  
of variables $\xi=x-k_1 t$, that results in 
\begin{equation}  
\label{caso11} 
{\partial \over \partial t} [\hat{p}(\xi,t)]^{\mu} =  
D~e^{\alpha(1-\nu/\mu) t} {\partial^2 \over \partial \xi^2} [\hat{p}(\xi,t)]^{\nu}. 
\end{equation} 
Now we change the time variable according to   
\begin{eqnarray} 
\hat{p}(\xi,t) &\leftrightarrow& \hat{p}(\xi,z(t)) \Longrightarrow  
{\partial \over \partial t} = \dot{z}(t) {\partial \over \partial z} \\ 
z(t) &=& \int_{0}^{t} e^{\alpha(1-\nu/\mu) \tau} d\tau  
     = {1 - e^{-\gamma t} \over \gamma} \qquad t \geq 0, 
\end{eqnarray} 
with $\gamma=-\alpha(\mu-\nu)/\mu$, and obtain the following equation valid for  
all $t \geq 0$  
\begin{equation} 
\label{caso12} 
{\partial \over \partial z} [\hat{p}(\xi,z)]^{\mu} =  
D~{\partial^2 \over \partial \xi^2} [\hat{p}(\xi,z)]^{\nu}.  
\end{equation} 
Hence, if $p_q^{(0)}(x,t)$ is the solution of Eq. (\ref{simple}),   
the solution with $F(x)=k_1$ plus absorption results to be  
\begin{equation} 
\label{caso13} 
p_q(x,t) = e^{-{\alpha \over \mu} t} p_q^{(0)}(x-k_1 t,z(t)).  
\end{equation} 
It is easy to check that this solution has the right limits for  
$\alpha \to 0$ and for $\alpha > 0$ and $\mu = \nu = 1$, i.e the standard 
Fokker-Planck equation (plus absorption).  
 
The new variable $z(t)$ plays the role of an {\it effective time}  
for the dispersion process. It exhibits markedly different behaviors 
depending on the ratio $\mu/\nu$. The case $\mu/\nu > 1$ $ (\gamma <0)$ 
corresponds to {\it superdiffusive} transport when there is no 
absorption \cite{inic1}. When absorption is present this {\it superdiffusion} 
is enhanced since the effective time $z$ grows exponentially as a function of 
the real time $t$. 
 
In the case $\mu/\nu <1$ $(\gamma > 0)$, which leads to {\it subdiffusion} 
for $\alpha = 0$, the presence of absorption also plays a key role. The 
effective time $z(t)$ converges to an asymptotic  
value: $\lim_{t \rightarrow \infty} z(t) = z_\infty = 1/\gamma$. Therefore,  
the distribution $p^{(0)}_q(\xi,z)$ evolves toward an asymptotic  
curve $p_q^{(0)}(\xi,z_\infty)$. Let us note that $z_\infty$ diverges  
whenever $\alpha \to 0$  or $\mu/\nu \to 1$. In Fig. \ref{compara-temporal}  
we compare, in the $\mu/\nu<1$ case, the time evolution of the distributions  
for $\alpha \neq 0$ and $\alpha =0$. We also compare these distributions  
with the shape of the $\alpha \neq 0$ asymptotic curve. For completeness, we 
include in Fig.\ref{z} the behavior of $z$ on $t$, also illustrating its  
dependence on $\mu/\nu$.  
Finally in the case $\mu=\nu$ ({\it normal diffusion}$+${\it absorption} for  
the quantity $\Phi(x,t)=\left[ p(x,t) \right]^{\mu}$) the change of variables  
becomes a simple time scaling.  
 
As already mentioned, Eq. (\ref{completa}) can be viewed as a classical 
diffusion equation for $\Phi(x,t)=\left[ p(x,t) \right]^{\mu}$, where the 
diffusivity depends on $\Phi(x,t)$  through $D(\Phi)=D \Phi^{\nu/\mu-1}$.  
Therefore, it becomes clear that the absorption can enhance (reduce) the  
diffusive transport whenever $\mu/\nu>1$ ($\mu/\nu <1$), namely: 
as absorption proceeds $\Phi$ decreases, yielding to an increase or not in  
$D(\Phi)$ according to the $\mu/\nu$ ratio. This qualitative description seems  
to be in complete agreement with the previous quantitative results.

\section{Solution for a linear force case} 
\label{sec-general} 
 
We now consider the case of a linear force, given by $F(x)=k_1 - k_2~x$ (here  
and in what follows we assume that $k_2 > 0$), whose general solution without  
absorption was found in \cite{inic1}. To start with, we assume that $\mu = \mu '$.  
With the hint of the change of variables made in the previous section  
we propose the following changes, that define the new variables 
\begin{eqnarray} 
\label{decaimiento} 
p(x,t) &=& e^{-w t} \hat{p}(\xi,z(t)) \\ 
\xi &=& x~g(t) + f(t),  
\end{eqnarray} 
with $w, ~g(t)$ and $f(t)$ to be determined. In terms of these new variables  
the time and space derivatives becomes 
\begin{eqnarray} 
{\partial \over \partial t} &=& {\partial \over \partial t} +  
\left( x~\dot{g}(t)+ \dot{f}(t) \right) {\partial \over \partial \xi}  
+ \dot{z}(t) {\partial \over \partial z} \\ 
{\partial \over \partial x} &=& g(t) {\partial \over \partial \xi}.  
\end{eqnarray} 
 
Taking into account these results and the form proposed for the solution 
(given by Eq. (\ref{decaimiento})), each separate term of Eq. (\ref{completa}) becomes 
\begin{eqnarray} 
{\partial \over \partial t} [p(x,t)]^{\mu} &=& 
-w \mu e^{-w \mu t} [\hat{p}(\xi,z)]^{\mu} + \\  
\nonumber 
&+& 
e^{-w \mu t}    
(x~\dot{g}(t)+\dot{f}(t)) {\partial \over \partial \xi} [\hat{p}(\xi,z)]^{\mu} \\  
\nonumber 
&+& e^{-w \mu t}    
 \dot{z}(t) {\partial \over \partial z} [\hat{p}(\xi,z)]^{\mu},  
\end{eqnarray} 
\begin{eqnarray} 
-{\partial \over \partial x} \left\{(k_1-k_2 x)[p(x,t)]^{\mu}\right\} &=&  
 k_2 [p(x,t)]^{\mu} - (k_1-k_2 x) {\partial \over \partial x}  
[p(x,t)]^{\mu}\\ 
\nonumber 
&=& e^{-w \mu t} \left\{ k_2 [\hat{p}(\xi,z)]^{\mu} -  
(k_1-k_2 x) g(t) {\partial \over \partial \xi} [\hat{p}(\xi,z)]^{\mu} \right\},  
\end{eqnarray} 
\begin{equation} 
D {\partial^2 \over \partial x^2} [p(x,t)]^{\nu} =  
D g^2(t) e^{-w \nu t} {\partial^2 \over \partial \xi^2} [\hat{p}(\xi,z)]^{\nu}.  
\end{equation} 
In this way, the equation we obtain for $\hat{p}(\xi,z)$ replacing  
into Eq. (\ref{completa}) with $F(x)=k_1 - k_2 x$ results (after arranging  
terms and multiplying by $e^{w \mu t}$) 
\begin{eqnarray} 
\dot{z}(t) {\partial \over \partial z} [\hat{p}(\xi,z)]^{\mu} &=&  
[w \mu + k_2 - \alpha] [\hat{p}(\xi,z)]^{\mu} \\  
\nonumber 
&& - [ (k_1-k_2 x) g(t) + x~\dot{g}(t)+ \dot{f}(t)]   
{\partial \over \partial \xi} [\hat{p}(\xi,z)]^{\mu} \\ 
\nonumber 
&& 
+ D g^2(t) e^{w (\mu - \nu) t} {\partial^2 \over \partial \xi^2} [\hat{p}(\xi,z)]^{\nu}.  
\end{eqnarray} 
In order to reduce the last equation to a one with a form similar to  
Eq. (\ref{simple}), we need to cancel the first two terms on  
the rhs, and reduce the coefficient of the third one to a constant.  
To operate with the second term, we shall cancel it for all values of $x$.  
These conditions yield the following equations 
\begin{eqnarray} 
0 &=& w \mu + k_2 - \alpha \\ 
0 &=& - k_2 g(t) + \dot{g}(t) \\ 
0 &=& k_1 g(t) + \dot{f}(t) \\ 
1 &=& g^2(t) e^{w (\mu - \nu) t} \dot{z}(t)^{-1},  
\end{eqnarray} 
rendering 
\begin{eqnarray} 
w &=& (- k_2 + \alpha) \mu^{-1} \\ 
g(t) &=& G e^{k_2t} \\ 
f(t) &=& H - G {k_1 \over k_2} e^{k_2 t} \\ 
z(t)-z(0) &=& G^2 \left\{ 1-e^{-\gamma t} \right\} \gamma^{-1},  
\end{eqnarray} 
with $\gamma = \left(- k_2(\mu + \nu) + \alpha (\nu - \mu) \right) \mu^{-1}$.  
In the general case, the values of the constants shall be chosen to fulfill some  
particular initial condition. Here, to simplify, we choose $G=1$, implying that  
we do not change the $x$ scale at $t=0$. Also, in order to make the change of  
space variables in such a way to have it centered at the potential minimum  
($\xi'=(x-{k_1 \over k_2})$) we adopt $H=0$. Finally we choose $z(0)=0$ to  
preserve the time origin. With these values we have  
\begin{eqnarray} 
\xi &=& \left( x - {k_1 \over k_2} \right)~ e^{k_2 t} \\ 
z(t)&=& \left\{ 1-e^{-\gamma t} \right\} \gamma^{-1},  
\end{eqnarray} 
and the solution of Eq. (\ref{completa}) with $F(x)=k_1 - k_2~x$ is  
\begin{equation} 
\label{final} 
p_q(x,t) = e^{{k_2 - \alpha \over \mu} t} p_q^{(0)}(\xi,z).  
\end{equation} 
This is the final result for the present case ($\mu=\mu'$). It is trivial to check  
its validity in some limits, the most obvious one is to choose $\alpha = 0$,  
recovering the solution of Ref. \cite{inic1}. With $\mu=\nu=1$ and  
$\alpha \ne 0$ we recover the simple case of diffusion in a harmonic  
potential with absorption. Also, if we consider the case $\mu=\nu$ 
and $\alpha \ne 0$, it is immediate to obtain (remember that $\mu=\nu$  
gives $q=1$!) 
\begin{equation} 
p_1(x,t) = e^{- {\alpha \over \mu} t}  
\frac 
{e^ 
{-\frac{k_2}{2 \mu D}\frac{[x - {k_1 \over k_2} - x_0 e^{-k_2 t}]^2}  
{1 - e^{-2 k_2 t}}}} 
{[{2 \pi \mu D \over k_2} (1 - e^{-k_2 t})]^{1/2 \mu}}.  
\end{equation} 
This result becomes obvious after making the change  
$p_1(x,t)^{\mu} = \phi(x,t)$, reducing the problem to an effective one  
of diffusion in a harmonic potential with absorption for $\phi(x,t)$.  
Clearly, even though the solution has a Gaussian form (times a decaying  
exponential term), the width of the Gaussian factor behaves ``anomalously"  
as it differs from the one in the associated Ornstein-Uhlenbeck process  
\cite{GARD}.  
 
As in the constant-force case, absorption process markedly influences  
the time evolution of dispersion. A straightforward calculation yields the  
dispersion of the distribution in the present case (i.e.  $\mu=\mu'$)  
\begin{equation} 
\label{dispersion} 
\langle (x- \langle x \rangle)^2 \rangle = {1 \over \beta[z(t)]} e^{-2 k_2 t}. 
\end{equation} 
In the superdiffusive case ($\mu/\nu>1$) $\beta(z(t))$ becomes asymptotically 
exponential, namely,  
$\beta(z(t))  
\propto  
\exp\left({{-2}{k_2(\mu+\nu)+\alpha(\mu-\nu) \over \mu + \nu } t }\right)$.  
Replacing this result in Eq. (\ref{dispersion}) we obtain the long time  
behavior of the dispersion  
\begin{equation} 
\label{dispersion2} 
\langle (x- \langle x \rangle)^2 \rangle \propto  
e^{2 \alpha {(\mu-\nu) \over (\mu + \nu)} t}. 
\end{equation} 
Therefore, the {\it superdiffusive} transport enhanced by absorption yields  
an exponentially increasing dispersion even in an attractive potential.   
The {\it subdiffusive} case ($ \mu / \nu < 1$) presents two 
different situations. Although in both cases the dispersion decays 
exponentially, when the absorption rate is small 
($\gamma < 0, \alpha < k_2 (\nu + \mu) / (\nu - \mu)$), absorption is 
the rate controlling process for dispersion 
\begin{equation} 
\langle (x- \langle x \rangle)^2 \rangle \propto  
e^{-2 \alpha {(\nu-\mu) \over (\nu + \mu)} t}. 
\end{equation} 
In the other case, when the absorption rate is large ($\gamma > 0$), 
the attractive force becomes rate limiting for dispersion process  
\begin{equation} 
\langle (x- \langle x \rangle)^2 \rangle \propto e^{-2 k_2 t}. 
\end{equation} 
 
In order to compare the influence of the absorption term on the solutions we  
have found, in Fig. \ref{alfa} we depict the solution given in Eq. (\ref{final}),  
in the case $\mu/\nu > 1$ (superdiffusion), for $\alpha=0$ and $\alpha \neq 0$  
at different times. In Fig. \ref{anomalo-normal}  we compare, in the subdiffusive  
case, the solution in Eq. (\ref{final}) when absorption is the rate limiting 
process for dispersion to the case when the attractive force controls 
dispersion. In both figures the differences between the characteristics of the  
different situations are apparent.  
 
\section{General Absorption Term} 
\label{sec-mu} 
 
In this section we consider Eq. (\ref{completa}), in the general case  
$\mu \neq \mu'$. The following ``simple" kinetic equation 
\begin{equation}  
\label{kinetic} 
{\partial \over \partial t} [p(t)]^\mu = -\alpha \left[ p(t) \right] ^{\mu'}, 
\end{equation} 
whose solution is 
\begin{equation}  
\label{exp_q}  
p(t) = [1-(1-q') \,{\alpha \over \mu} \, t]^{-{1 \over {(1-q')}}},   
\end{equation}  
where $q'=1-\mu'+\mu$, strongly suggests to replace the exponential 
in the change of variables in Eqs. (\ref{caso13}) and (\ref{decaimiento}), by  
the {\it q'-exponential} function defined by Eq. (\ref{exp_q}) \cite{borges}.  
The ordinary exponential function is recovered when $q' \to 1$. 
If we try this possibility, together with the Ansatz in Eq. (\ref{sol1c}), 
it immediately leads to the condition $\mu=\mu'$. This result becomes 
apparent when analyzing Eq. (\ref{general}) in terms of  
$\Phi(x,t)=\left[ p(x,t) \right] ^{\mu}$. 
The form proposed in Eq. (\ref{reduction}) allows to reduce the general equation, 
eliminating the absorption term, only when absorption is proportional 
to $\Phi(x,t)$. However, the general case with $\mu \neq \mu'$ will have a solution 
whose scaling properties can be determined. 
 
To find such scaling behavior it is enough to consider the simplified situation without  
external force, that is 
\begin{equation} 
\label{simplified} 
{\partial \over \partial t} [p(x,t)]^{\mu} =  
D {\partial^2 \over \partial x^2} [p(x,t)]^{\nu} - \alpha [p(x,t)]^{\mu'}.  
\end{equation} 
We consider the following Ansatz 
\begin{equation} 
p(x,t) = \varphi(t) \Theta(\xi),  
\end{equation} 
with $\xi = \psi(t)~x$. Replacing this into Eq. (\ref{simplified}), we obtain the 
functions $\varphi(t)$ and $\psi(t)$ as  
\begin{eqnarray} 
\varphi(t) &=& \left[ 1 + (1 - q')~(\alpha/\mu) \,t \right]^{-{1 \over {1-q'}}} \\   
\psi(t) &=& \left[ 1 + (1 - q')~(\alpha/\mu) \,t \right]^{-{1 \over 2} {\mu'- \nu \over (1-q')}}  
\end{eqnarray}  
where $q' = 1 -\mu' + \mu$.  Hence, Eq. (\ref{simplified}) for $\Theta(\xi)$  
reduces to an ordinary differential equation on the variable $\xi$ 
\begin{equation}  
\label{theta}  
\Theta^{\mu}+\left[ {\mu' - \nu \over 2 \mu }\right]\, \xi \, {d \over d \xi} \Theta^{\mu} = 
D {d^2 \over d\xi^2} \Theta^{\nu}+\Theta^{\mu'}  
\end{equation}  
  
Once again, the previous results can be interpreted in terms of $\Phi(x,t)$. 
Writing the absorption term as $-(\alpha \Phi^{\mu'/\mu-1}) \Phi$, it can 
be seen that in the case $\mu'<\mu$ the absorption process is enhanced as 
$\Phi$ decreases with time. This leads to a finite time 
$t_{c}=\mu/\alpha(\mu-\mu')$, where $\Phi$ becomes zero. 
On the other hand, when $\mu'> \mu$ we can obtain the asymptotic dispersion,  
even though the ordinary differential equation for $\Theta(\xi)$, 
Eq. (\ref{theta}), is too complicated to be solved analytically.   
However, for the n{\it th}-moment of the distribution, we obtain 
\begin{eqnarray} 
\langle x^{2n} \rangle  
&=& \left[ \int ~dx ~x^{2n} ~p(x,t) ~\right] \left/ \left[ ~\int ~dx~ p(x,t) ~\right] 
\right. \\ \nonumber 
&=& \left[ \int ~dx ~x^{2n} ~\varphi(t)~ \Theta(\psi(t)~x) ~\right] \left/  \left[ \int ~dx ~\varphi(t) 
~\Theta(\psi(t)~x) ~\right] \right.\\ \nonumber 
&=& \psi(t)^{-2n} ~\left[ ~ \int ~d\xi ~ \xi^{2n} ~ \Theta(\xi) ~ \right] \left/  
\left[ ~ \int d\xi ~\Theta(\xi) ~ \right] \right.\\ \nonumber 
&=& \psi(t)^{-2n} A_{2n} \\ 
\langle x^{2n+1} \rangle &=& 0,  
\end{eqnarray} 
yielding  
\begin{equation} 
\langle \left( x - \langle x \rangle \right)^2 \rangle = \langle x^{2} \rangle = 
\psi(t)^{-2} A_{2} \sim t^{ \left( {\mu' - \nu \over (1-q')} \right)} =  
t^{ \left( {\mu' - \nu \over \mu' - \mu } \right)}. 
\end{equation} 
Hence, it is clear that, as in the previous case ($\mu'=\mu$), $\mu/\nu<1$  
corresponds to subdiffusion whereas the case $\mu/\nu>1$ corresponds to  
superdiffusive transport.

\section{Final Remarks} 
 
The Fokker--Planck equation that was generalized to a nonextensive  
scenario \cite{inic3} has been  
further generalized to include the possibility of an absorption process.  
We have shown that the exact solutions of Eq. (\ref{completa})  
(a nonlinear Fokker--Planck equation subject to linear forces)  
found in \cite{inic1} when $\alpha = 0$, can be extended for the case  
$\alpha \neq 0$ and $\mu ' = \mu$. However, in the general case  
$\mu ' \neq \mu$, we have been only able to obtain the scaling properties  
of the solution (whose analytical form we cannot obtain),  
and the asymptotic behavior of the whole hierarchy of moments.  
 
Summarizing our results for the nonlinear process of anomalous diffusion 
plus absorption, as described by Eq.(\ref{completa}), we have found 
that the solution \\
{\bf -} in a constant force field and for $\mu/\nu > 1$ shows a 
superdiffusive behaviour that is {\it enhanced} when $\alpha \neq 0$
($\gamma < 0$), \\
{\bf -} also, when $\gamma > 0$ ($\mu/\nu < 1$) the concentration 
reaches an asymptotic constant profile, \\
{\bf -} for a linear force and $\gamma < 0$, we find an exponentially 
increasing dispersion for superdiffusion, \\
{\bf -} also, in the linear force case, an exponentially decreasing
dispersion arises for subdiffusion (gamma > 0),  where absorption is the rate
controlling process for dispersion when absorption rate is small
($\alpha < k_2 (\nu + \mu) / (\nu - \mu)$), while the attractive force becomes 
the rate limiting dispersion process when absorption is large enough 
($\alpha > k_2 (\nu + \mu) / (\nu - \mu)$).

The present results gives further support to the argument \cite{inic1} that 
a generalized thermostatistics including nonextensivity constitutes an adequate  
framework  within which it is possible to  
unify both normal and correlated anomalous diffusion \cite{inic1},  
extended now to the case when an absorption process is also present.  
Also, as indicated in \cite{inic4}, this kind of work points out the convenience  
of paying more attention to the thermodynamic aspects of non--Fickian diffusion.    
Moreover, it has been suggested \cite{inic1} that even Levy-like anomalous  
diffusion \cite{Levy} (that can be discussed by means of linear Fokker--Planck  
equations with fractional derivatives) can be included within the present common  
framework of nonlinear Fokker--Planck equations with fractional derivatives.  
 
This work also opens the possibility of analyzing reaction--diffusion systems on a  
fractal substratum, by considering nonlinear Fokker--Planck equation with other  
forms of reaction terms. This problem will be the subject of further work.  
 
\vskip.5cm 
 
\noindent {\bf Acknowledgments:} GD thanks for support from UBA and CONICET.  
HSW acknowledges financial support from CONICET and ANPCyT (Argentinian agencies).  
CT acknowledges the warm hospitality extended to him during his stay at Instituto  
Balseiro, and partial support from CNPq, FAPERJ and PRONEX (Brazilian agency).

\begin{figure}[h] 
\caption[pt]{The evolution of the distribution $p_q^{(0)}(\xi,z(t))$, 
as given in Eq. (\ref{caso13}), in 
the case $\mu/\nu = 2/3 < 1$. It is shown at times $t =0.2,\, 2.5$ and $5$ 
for $\alpha =0$ and $\alpha =1$ ($\mu=1, \nu=1.5, D=1, k_1=5, q=0.5$). The 
dashed line corresponds to the $\alpha=1$ asymptotic distribution shape.} 
\label{compara-temporal} 
\end{figure} 
 
\begin{figure}[h] 
\caption[zt]{Time dependence of the {\it effective time} $z$ for several  
values of $\gamma$. It is apparent that $z(t)$ saturates at a finite value 
as $t \rightarrow \infty$ ($z_\infty = 1/\gamma$) whenever  
$\gamma > 0$. In the case $\gamma=0$ the effective time becomes equal to the  
real time: $z=t$. For $\gamma < 0$ the effective time  
grows exponentially with $t$}  
\label{z} 
\end{figure} 
 
\begin{figure}[h] 
\caption[alfa]{Evolution of the distribution $p_q^{(0)}(\xi,z(t))$,  
as given in Eq. (\ref{final}), in the case $\mu/\nu = 2 > 1$. It is shown 
at times $t = 0.5, 0.8$ and $1.0$ for $\alpha =0$ and $\alpha =5$  
($\mu=1, \nu=0.5, D=1, k_1=0, k_2=1.0, q=1.5$). The dotted lines correspond   
to the $\alpha=0$ distribution, while the solid lines correspond to the 
$\alpha=5$ case.} 
\label{alfa} 
\end{figure} 
 
\begin{figure}[h] 
\caption[a-n]{Evolution of $p_q^{(0)}(\xi,z(t))$, as given by Eq. (\ref{final}), in 
the case $\mu/\nu = 2/3 < 1$. We show it at times $t =0.2, 1$ and $5$ 
for $\alpha =0.5$ and $\alpha =2.0$ ($\mu=1, \nu=1.5, D=1, k_1=0,  
k_2=0.2, q=0.5$). The dashed lines correspond to the case $\alpha=0.5$ where 
the attractive force becomes dispersion rate limiting. The solid lines correspond  
to $\alpha=2.0$ where absorption is the rate controlling process for 
dispersion.}  
\label{anomalo-normal} 
\end{figure} 
 
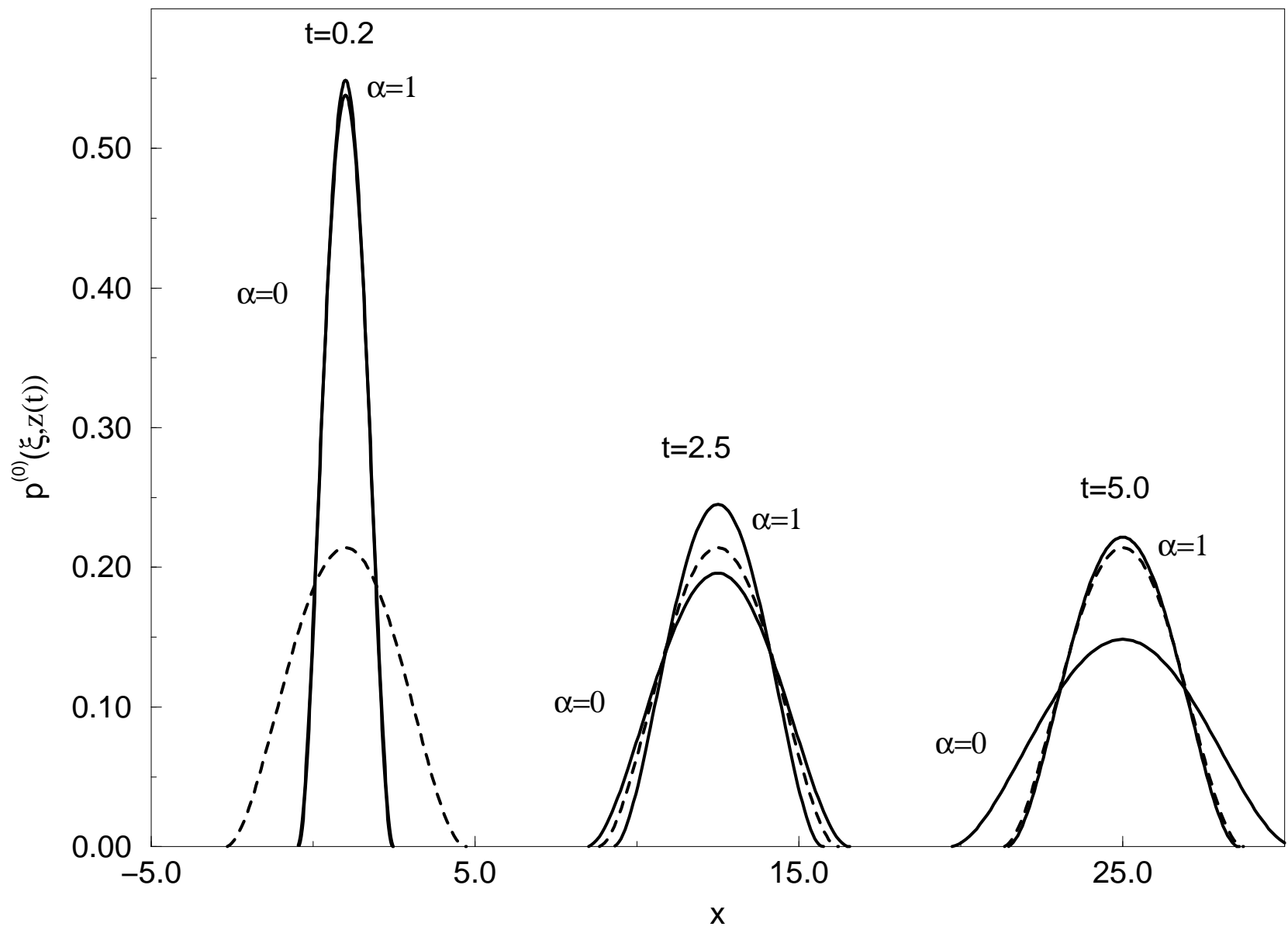
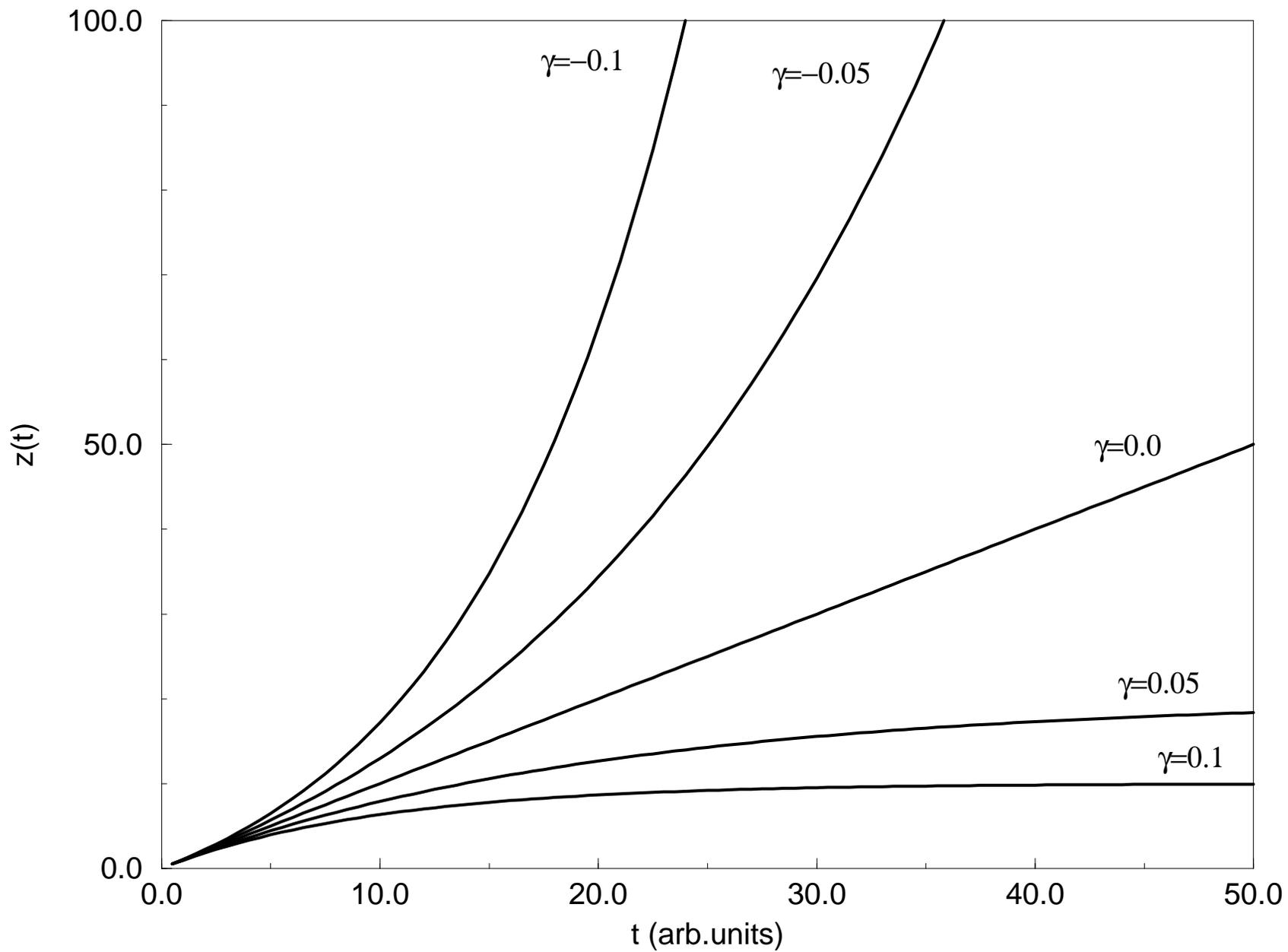
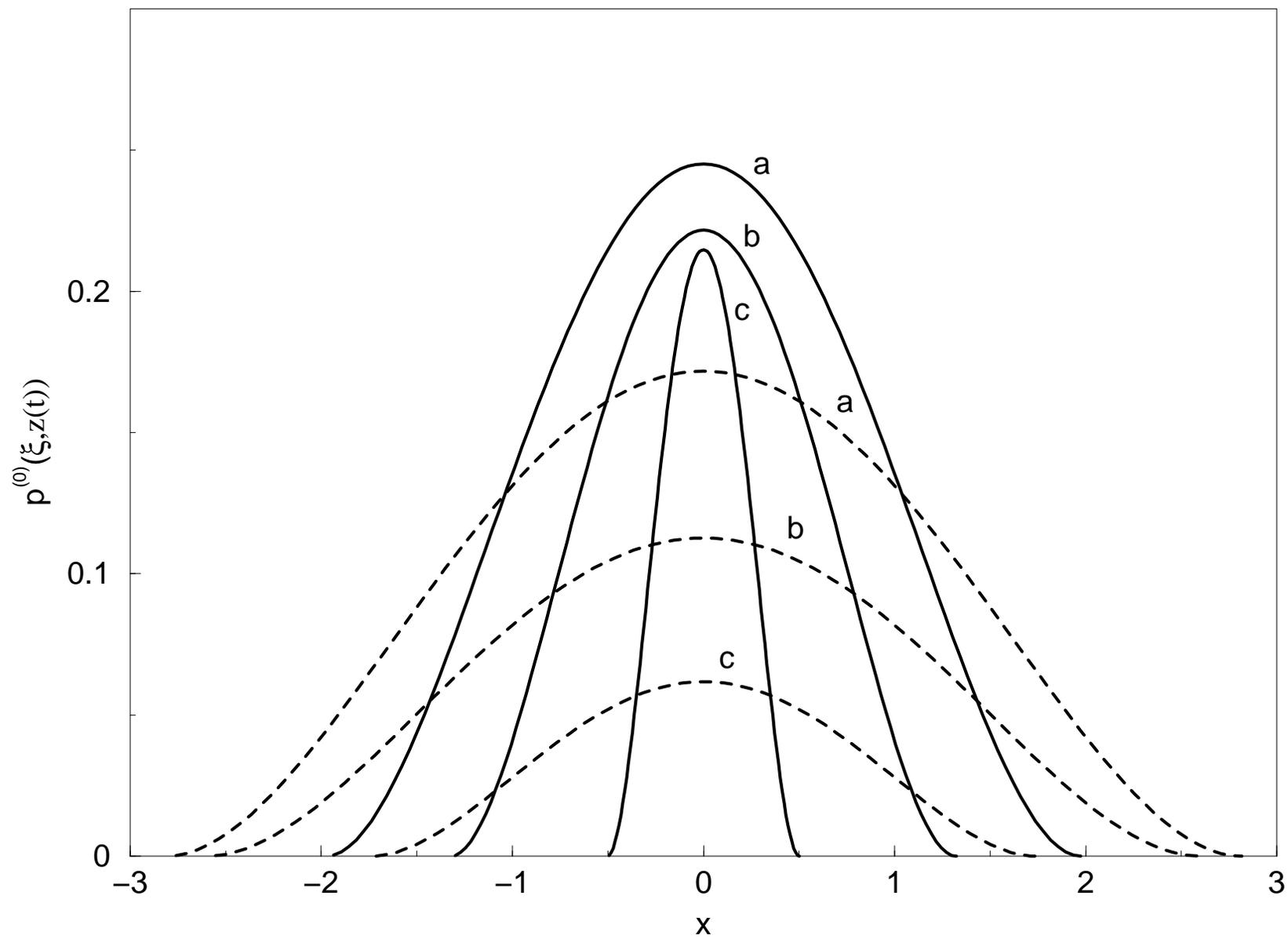
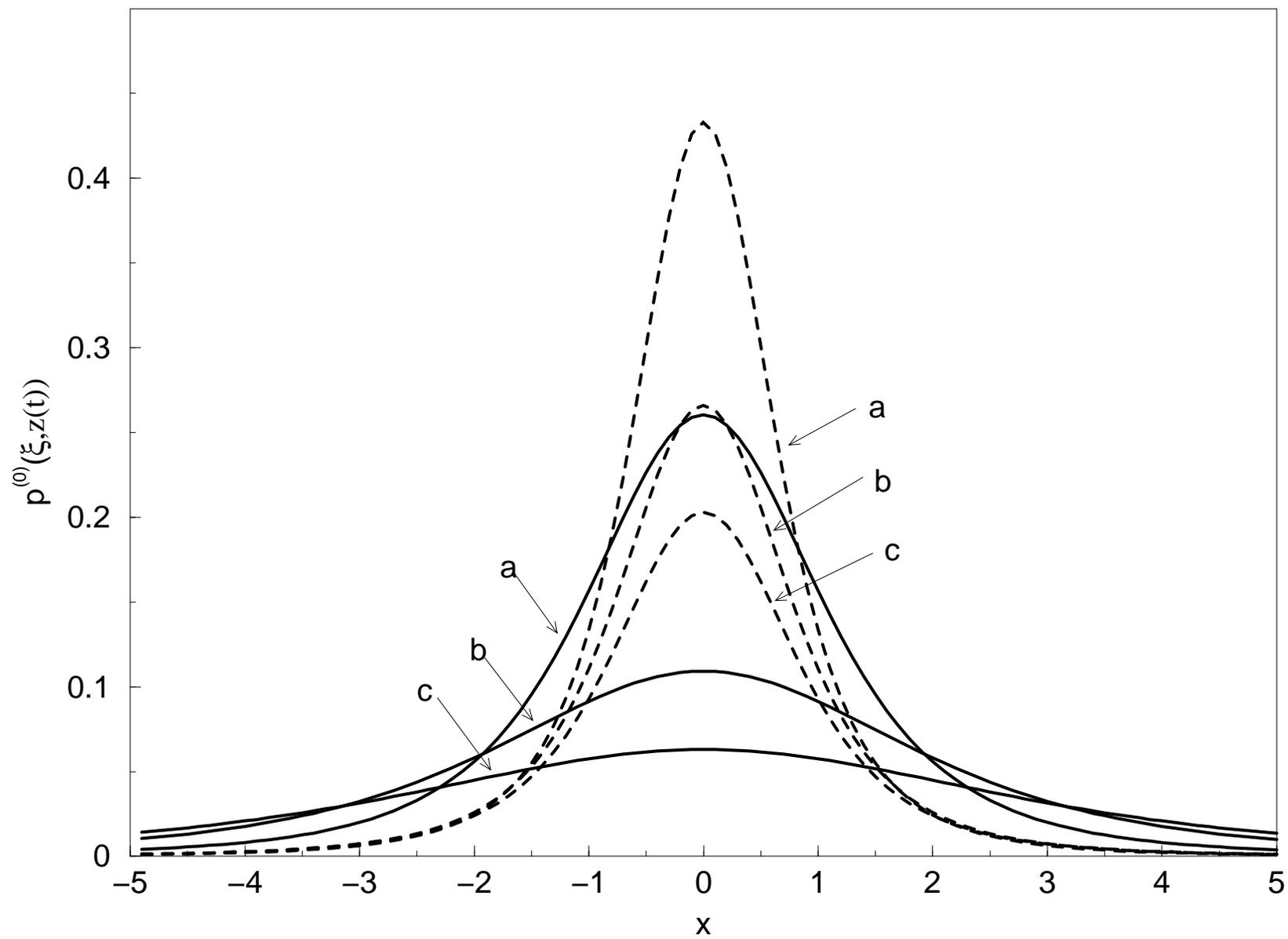
\end{document}